\documentclass[english]{article}
\usepackage[T1]{fontenc}
\usepackage[latin9]{inputenc}
\usepackage{amsmath}
\usepackage{amssymb}
\usepackage{graphicx}

\makeatletter

\providecommand{\tabularnewline}{\\}

\usepackage{amsfonts}

\makeatother

\usepackage{babel}
\begin{document}

\title{Model for the shear viscosity of suspensions of star polymers and
other soft particles}

\author{Carlos I. Mendoza%
\thanks{E-mail: cmendoza@iim.unam.mx%
}\\
Instituto de Investigaciones en Materiales, Universidad Nacional\\
Autónoma de México, Apdo. Postal 70-360, 04510 México, D.F., Mexico}
\maketitle
\begin{abstract}
We propose a model to describe the concentration dependence of the
viscosity of soft particles. We incorporate in a very simple way the
softness of the particles into expressions originally developed for
rigid spheres. This is done by introducing a concentration-dependent
critical packing, which is the packing at which the suspension looses
fluidity. The resultant expression reproduces with high accuracy the
experimental results for suspensions of star polymers in good solvents.
The model allows to explain a weak increase of the viscosity observed
in the case of diblock copolymer stars suggesting that the reason
for this peculiar behavior is mainly a consequence of the softness
of the particles. In the semi-dilute regime, suspensions of star polymers
are modeled using the Daoud-Cotton picture to complete the description
in the whole concentration regime.
\end{abstract}

It is of basic and applied importance to understand the flow behavior
of fluids containing colloidal particles and macromolecules. The range
of industrial applications of such complex fluids range from motor
oils, coatings, drilling fluids and food products among others. Many
colloidal suspensions consist of hard particles whose rheological
properties have been studied profusely. By contrast, much less attention
has been paid to suspensions containing soft colloids. Representatives
of soft colloids include deformable particles and particles that in
addition may interpenetrate to some extent. 

The interest in studying soft particles stems from the desire to explore
the behavior of a number of complex fluids composed of nonrigid structures
such as polymerically stabilized colloidal spheres \cite{vlassopoulos},
block copolymer micelles \cite{gast}, star polymers \cite{semenov},
hard spheres with a grafted polymer brush \cite{castaing}, dendritically
branched polymers \cite{asteriadi} and others. In particular, star
polymers which are branched macromolecules with a small central core
from where $f$ arms of identical linear polymer chains, each one
containing $N$ segments emerge are considered an important prototypical
model of soft particle because their size and architecture can be
tailored between that of linear polymers and ultra-soft colloids to
nearly hard-spheres \cite{fedosov}.

Pioneering theoretical work to calculate the viscosity of dilute colloidal
suspensions was initiated by Einstein who obtained an expression to
calculate the viscosity as function of the concentration of dilute
suspensions of spherical hard particles \cite{einstein}. Various
models to calculate the viscosity of dilute suspensions of particles
other than solid spheres have been proposed. Among them we can mention
the case of emulsions of spherical droplets \cite{taylor}, homogeneously
porous rigid spheres \cite{brinkman1}-\cite{ohshima0} and uncharged
spherical soft particles \cite{ohshima1}.

For non-dilute systems, the suspension rheology is determined by the
interplay between the direct particle-particle interactions and the
solvent-mediated hydrodynamic interactions (HIs) \cite{abade}. Their
many body nature impose a formidable difficulty for the calculation
of rheological quantities such as the shear viscosity of the system.
The fact that many particles are soft complicates even more the calculation.
Nonetheless, it is a common practice to correlate the concentration
dependence of the viscosity of star polymers and other hairy particles
using phenomenological models designed for solid hard spheres \cite{jones},
\cite{roovers}. In particular, experimental data are frequently compared
with the expression proposed by Krieger and Dougherty for hard colloidal
particles \cite{krieger}

\begin{equation}
\eta\left(\phi\right)=\eta_{0}\left[1-\left(\frac{\phi}{\phi_{\max}}\right)\right]^{-\left[\eta\right]\phi_{\max}},\label{krieger}
\end{equation}
where $\eta\left(\phi\right)$ is the viscosity of the suspension,
$\eta_{0}$ the viscosity of the background solvent, $\phi$ is the
volume fraction of the colloidal particles, $\phi_{\max}$ is the
volume fraction at maximum packing and $\left[\eta\right]=5/2$ is
the intrinsic viscosity of hard spheres. Comparing experimental data
for star polymers with this expression might be accurate for stars
with a large number of short arms since they behave essentially as
hard spheres. However, such comparisons do not give satisfactory results
in the case of stars with larger arms because in this case the particles
are much softer. In addition to this, it has been shown that the Krieger
and Dougherty expression underestimates the viscosity of the suspension
at large volume fractions even in the case of hard spheres, as explained
thoroughly in Ref. \cite{santamaria-holek}.

Alternative expressions have been devised to predict the viscosity
for concentrated suspensions including in a more accurate way hydrodynamic
and excluded volume interactions. Using a differential effective medium
technique based on a progressive addition of particles to the sample
in which the new particles interact in an effective way with those
added in previous stages, we have shown \cite{santamaria-holek}-\cite{mendoza3}
that the static and high-frequency viscosities of suspensions of colloidal
particles can be correlated with the expression

\begin{equation}
\eta\left(\phi\right)=\eta_{0}\left(1-\frac{\phi}{1-k\phi}\right)^{-\left[\eta\right]},\label{viscosity2}
\end{equation}
where $\left[\eta\right]$ is the intrinsic viscosity and $k$ a fitting
constant related to the critical volume fraction $\phi_{c}$ at which
the suspension loses its fluidity and is given by
\begin{equation}
k=\frac{1-\phi_{c}}{\phi_{c}}.\label{k}
\end{equation}
Expression (\ref{viscosity2}) incorporates crowding effects and hydrodynamic
interactions \cite{santamaria-holek}-\cite{mendoza3} and has been
applied successfully to suspensions of hard spheres \cite{mendoza},
emulsions of spherical droplets \cite{mendoza2}, suspensions of arbitrarily-shaped
hard particles \cite{santamaria-holek}, suspensions of rigid core-shell
permeable particles \cite{mendoza3} and suspensions with power-law
matrices \cite{tanner} with excellent results. However, as in the
case of the Krieger and Dougherty model, when applied to soft (interpenetrable)
particles, Eq. (\ref{viscosity2}) fails to reproduce the viscosity-concentration
curves satisfactorily.

The purpose of the present work consists in propose a method to incorporate
the softness of the particles in the above mentioned expressions to
obtain a model that better correlates with experimental results.

\begin{figure}[ptb]
\centering{}\includegraphics{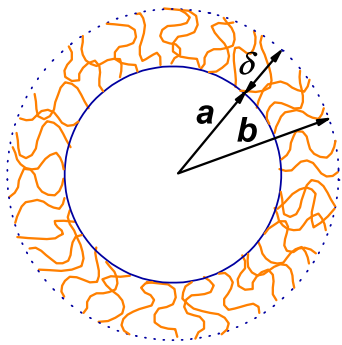}\caption{Schematics of a model for a suspension of core-shell spheres. A hard
core of radius $a$ covered by a porous layer of thickness $\delta$.
The total radius of the particle is $b=a+\delta$.}
\label{core-shell}
\end{figure}

Since we have in mind to apply the model to suspensions of star-branched
polymers we have to take into account the possibility that the particles
are permeable to some extent to the surrounding solvent. To do so,
we consider a core-shell particle consisting of a spherical hard core
of radius $a$ coated with a surface soft layer of thickness $\delta$.
The soft layer may represent a grafted polymer brush \cite{brader}
or the outer sections of a star polymer, as represented in Fig. \ref{core-shell}.
Thus, the polymer-coated particle or star polymer has a non-porous
inner radius $a$ and an a permeable outer shell of radius $b=a+\delta$.
The volume fraction $\phi$ is defined as the volume occupied by the
particles including the shell, divided by the volume of the sample.
To model the permeability of the outer shell, the polymer segments
are regarded as resistance centers distributed in the permeable polymer
layer, exerting frictional forces $-\gamma\mathbf{u}$ on the liquid
flowing in the layer, where $\mathbf{u}$ is the liquid velocity relative
to the particle and $\gamma$ is the frictional coefficient. The result
of this model is \cite{ohshima1} $\left[\eta\right]=(5/2)L_{2}\left(\lambda b,a/b\right)/L_{1}\left(\lambda b,a/b\right)$,
where $\lambda=\sqrt{\gamma/\eta_{0}}$. Expressions for $L_{1}\left(\lambda b,a/b\right)$
and $L_{2}\left(\lambda b,a/b\right)$ can be found in Ref. \cite{mendoza3}.
Therefore, the permeability of the outer shell is incorporated in
the intrinsic viscosity $\left[\eta\right]$ of the core-shell spheres.

If we consider that the outer core of the soft sphere do not exert
frictional forces on the liquid flowing through it, that is, $\lambda=0$,
then the expression for $\left[\eta\right]$ simplifies considerably
and one gets
\begin{equation}
\left[\eta\right]=\frac{5}{2}\left(\frac{a}{b}\right)^{3}.\label{ohshimalamdacero}
\end{equation}

The present model, Eq. (\ref{viscosity2}) together with Eq. (\ref{ohshimalamdacero})
allows to obtain the viscosity as a function of the concentration
of rigid porous core-shell particles and has been successfully applied
to calculate the static viscosity of surfactant coated particles and
the high-frequency viscosity of homogeneous porous particles \cite{mendoza3}.
However it does not give satisfactory results for suspensions of star
polymers.

As mentioned above, experimental results of the concentration-dependent
viscosity of star polymers can not be satisfactorily reproduced with
models developed for hard colloids like the Krieger-Dougherty model,
Eq. (\ref{krieger}) or the one given by Eq. (\ref{viscosity2}) even
when the intrinsic viscosity for core-shell particles, Eq. (\ref{ohshimalamdacero}),
is used. The problem with these expressions is that they assume that
the particles are rigid. It is apparent that the softness of the stars
must be taken into account to obtain better results. Here we propose
to model the softness of the particles by assuming that the critical
volume fraction, $\phi_{c}$, is larger than the random close packing
value $\phi_{RCP}$ since soft particles can interpenetrate each other.
Specifically, we propose the following \textit{ansatz}

\begin{equation}
\phi_{c}=\phi_{RCP}+\beta\phi^{\alpha},\label{softness}
\end{equation}
where $\alpha$ and $\beta$ are unknown constants that should be
determined from the fitting to the experimental data. Expression (\ref{softness})
can be interpreted as follows, when the volume fraction is very low,
the soft particles do not interact too much with each other and therefore
they do not overlap too much on average. In this situation they behave
as rigid impenetrable spheres and $\phi_{c}=\phi_{RCP}$. As the volume
fraction is increased, the particles are forced to collide more often
increasing the average overlap which in turn increases the value of
$\phi_{c}$. Eq. (\ref{softness}) then proposes that this increase
follows a power law on the volume fraction.

\begin{figure}[ptb]
\begin{centering}
\includegraphics[scale=0.75]{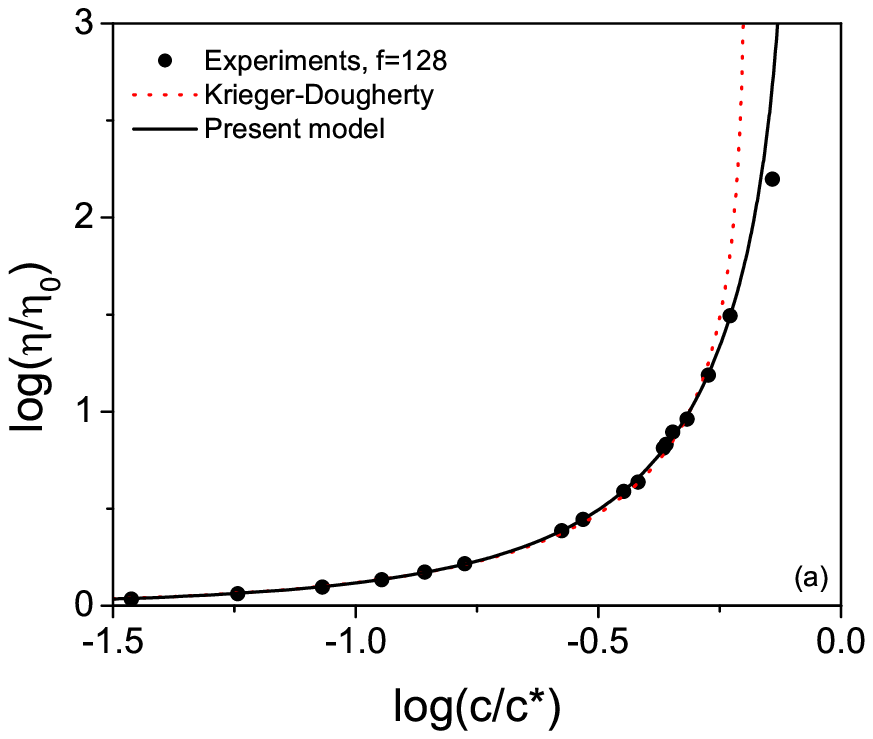}\includegraphics[scale=0.75]{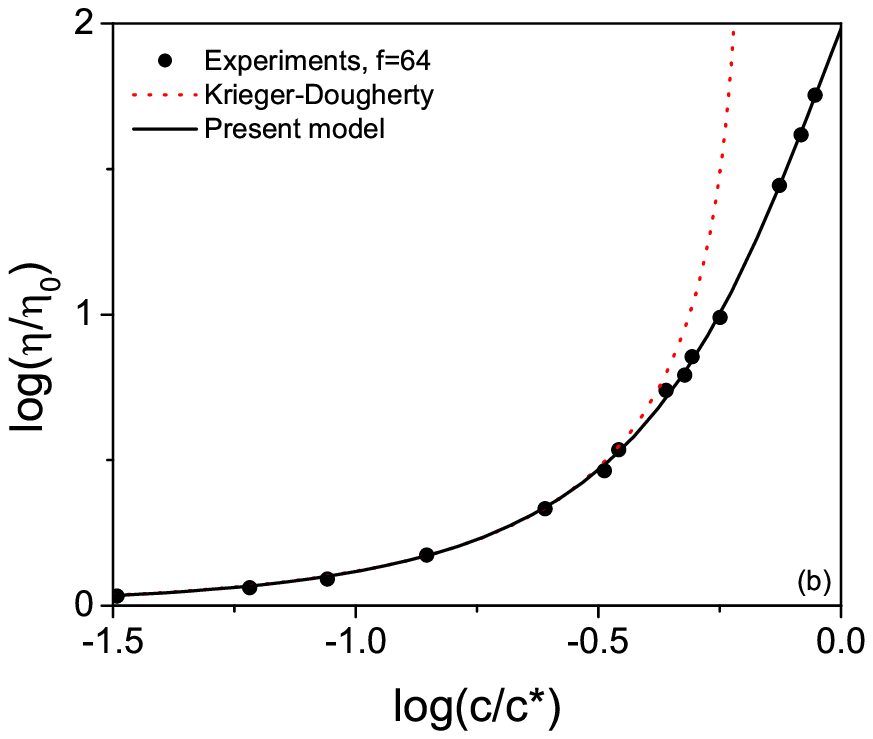}
\par\end{centering}

\centering{}\includegraphics[scale=0.75]{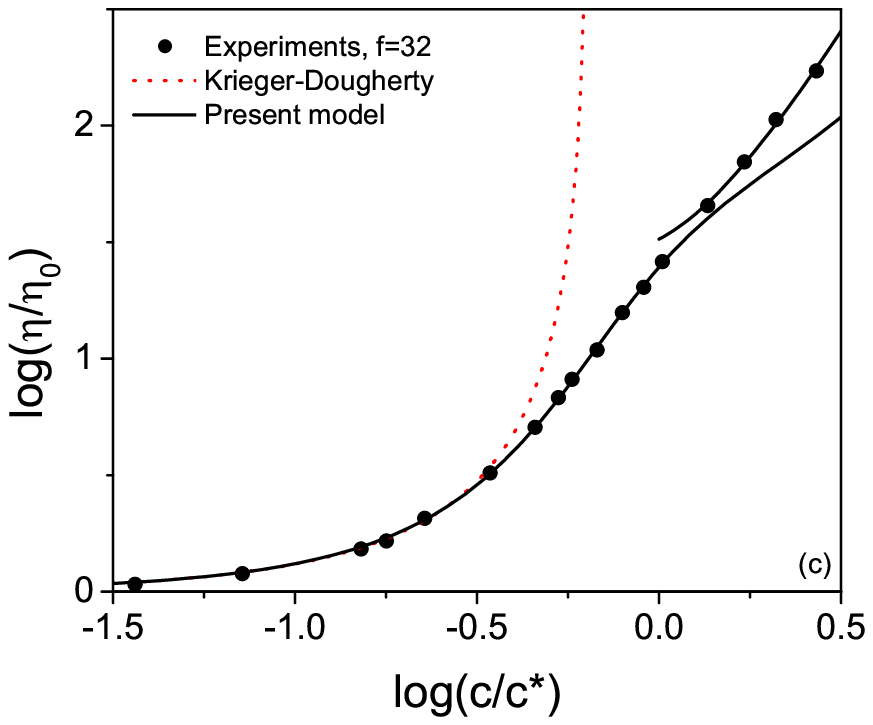}\includegraphics[scale=0.75]{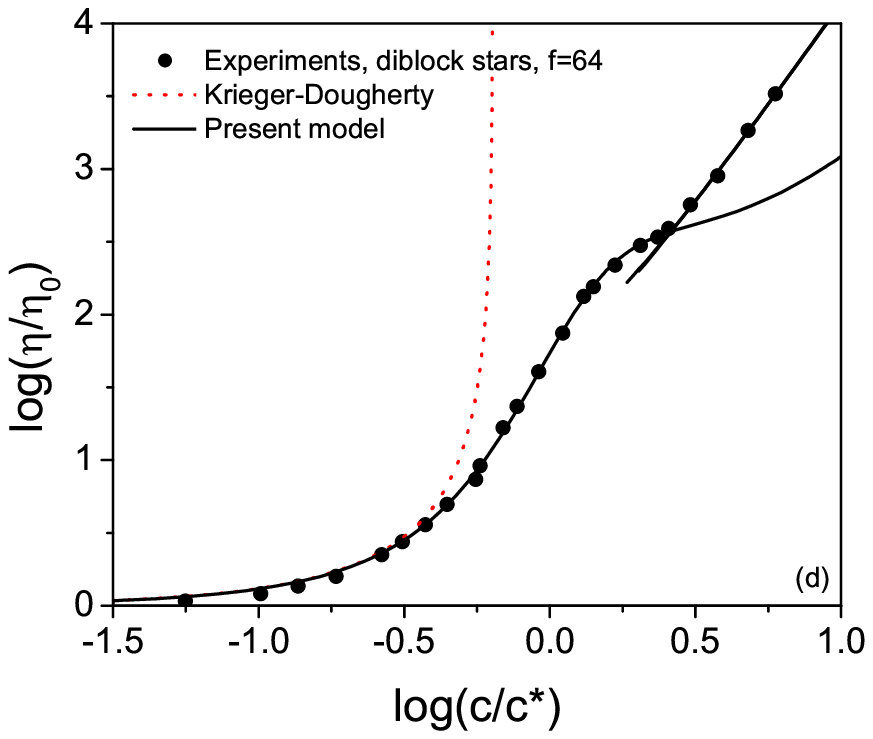}\caption{Relative viscosity $\eta(\phi)/\eta_{0}$ as a function of the particle
volume fraction $\phi$ for star polymers with (a) $128$, (b) $64$,
and (c) $32$ arms and (d) diblock copolymer micelle with $64$ arms
in the good solvent toluene. Experimental data taken from Ref. \cite{roovers}.}
\label{roovers1}
\end{figure}

To test the correctness of this proposal we compare the predictions
of our model given by Eq. (\ref{viscosity2}) including the possibility
that the outer sections of the star are solvent permeable, using the
intrinsic viscosity (\ref{ohshimalamdacero}). Finally and most importantly,
we incorporate the softness of the particles substituting Eq. (\ref{softness})
in Eq. (\ref{k}). This is done in Fig. \ref{roovers1} where we compare
the predictions of our model with the experimental results of Roovers,
Ref. \cite{roovers} for the relative viscosity of polybutadiene star
polymers in the good solvent toluene. Typically, the viscosity vs
concentration curves for star polymers are plotted in terms of the
polymer concentration $c/c^{*}$, where $c^{*}$ is the overlap concentration
defined by $c^{*}=\left[\left(N_{A}/M\right)\left(4\pi/3\right)R_{V}^{3}\right]^{-1},$
where $N_{A}$ is Avogadro's number, $M$ the molecular weight, and
$R_{V}$ the viscosimetric equivalent hard-sphere radius of the polymer
coil. On the other hand, models for the viscosity of colloidal suspensions
like Eq. (\ref{viscosity2}) are written in terms of the particle
volume fraction $\phi$. In the dilute regime these two quantities
are related by 
\begin{equation}
\phi=c/c^{*},\label{phistar1}
\end{equation}
which allows to plot our model in terms of the polymer concentration.
In Fig. \ref{roovers1} we perform such plot. Panel (a) shows the
experimental results for stars with $f=128$ arms. Notice that the
viscosity-concentration curve is not very different from the hard-sphere
result at low concentrations. However at large concentrations a slight
departure from the hard sphere behavior is observed. Our model reproduces
very well this departure with the fitting parameters shown in Table
\ref{tabla 2}. If the number of arms is decreased, one expects a
smaller steric repulsion between the arms of different stars allowing
a larger overlap and therefore a larger departure from the hard-sphere
behavior. This is shown in Fig. \ref{roovers1} (b) where results
for a star with $64$ arms are shown. The softness of the star is
larger then in the case of the star with $128$ arms and the departure
from the hard-sphere behavior is larger. Our model predicts correctly
the experimental data and the increased softness of this case is reflected
in a larger value of the parameter $\beta$, as shown in Table \ref{tabla 2}.
Panel (c) shows the results for a star with $32$ arms. The low number
of arms of this case allows a much larger overlap allowing that concentrations
above $c^{*}$ can be reached. Experimental data show clearly a different
behavior below and above $c^{*}$. The region below $c^{*}$ can still
be described with our model with an even larger value of $\beta$
than in the previous cases. However, above $c^{*}$, the overlap is
so large that the conformation of the star polymers should be described
differently. Here we use the Daoud and Cotton picture \cite{daoud}
to describe the star conformation in this semi-dilute regime.

\begin{figure}[ptb]
\centering{}\includegraphics[scale=0.75]{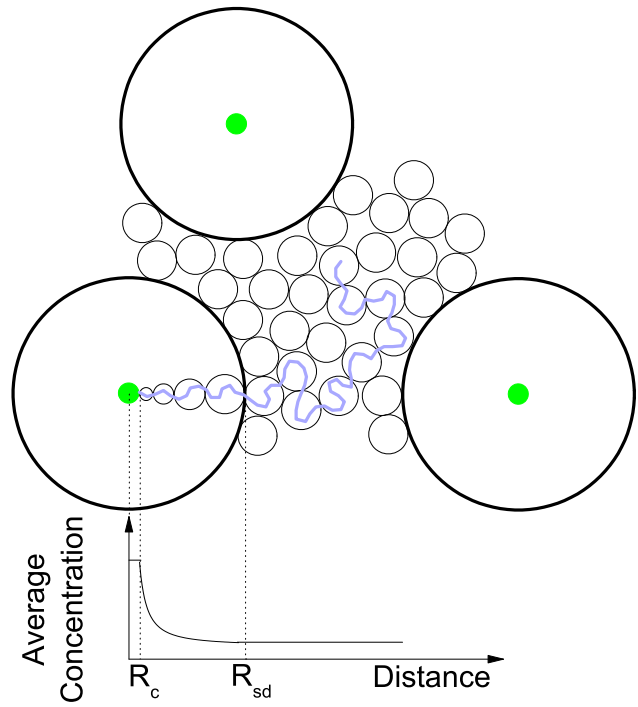}\caption{Daoud and Cotton model for a semi-dilute suspension of star polymers.}
\label{daoud-cotton}
\end{figure}

In the Daoud and Cotton model \cite{daoud}, a single star polymer
is regarded as a spherical region of a semi-dilute polymer solution
with a local, position dependent screening length $\xi\left(r\right)$,
where $r$ is the distance from the center. This is represented pictorially
by associating with each arm a string of blobs of increasing size
$\xi\left(r\right)$ as shown in Fig. \ref{daoud-cotton}. The blob
size varies as $\xi\left(r\right)\simeq rf^{-1/2}$ and the corresponding
local polymer volume fraction as $c_{s}\left(r\right)\simeq f^{2/3}\left(b/r\right)^{4/3}$.
The size of the star can be obtained from monomer conservation \cite{marques}
to obtain $R\simeq\beta\xi_{p}Nf^{1/5}$, where $\xi_{p}$ is the
Kuhn length. Above the overlapping concentration $c^{*}=fN\xi_{p}^{3}/R^{3}$,
the outer sections of the arms overlap and build a semi-dilute solution,
often called the sea of blobs, where the inner parts of the actual
stars are embedded and where the polymer concentration of the sea
of blobs levels off (see Fig. \ref{daoud-cotton}). The stars embedded
in the sea of blobs occupy a volume fraction $\phi$ and the sea of
blobs the remaining fraction $1-\phi$. Above the overlapping concentration
$c^{*}$, the values of the polymer concentration in the sea of blobs
$c_{p}$, of the fraction occupied by the stars $\phi$, and of the
radius of the embedded stars $R_{sd}$ can be calculated from mass
conservation \cite{marques}:

\begin{equation}
\phi=k_{1}\left(\frac{c^{*}}{c}\right)^{5/4},\label{phistar}
\end{equation}
\begin{equation}
c_{p}\simeq c,\label{phi sea of blobs}
\end{equation}
and
\begin{equation}
R_{sd}=Rk_{2}\left(\frac{c^{*}}{c}\right)^{3/4}.\label{radius inner spheres}
\end{equation}
The constant $k_{1}$ can be obtained from continuity of the volume
fraction at $\phi=\phi_{RCP}=0.637$ from Eqs. (\ref{phistar1}) and
(\ref{phistar}) to get $k_{1}=0.3625$, while the constant $k_{2}$
is obtained by continuity from Eqs, (\ref{phistar1}) and (\ref{radius inner spheres})
since $R_{sd}=R$ at the same point, to get $k_{2}=0.7130$. Notice
that an increase of the polymer concentration $c$ leads to an increase
of the fraction of space occupied by the sea of blobs and a shrinkage
of the inner star dimensions. Additionally, the stars are embedded
in a solvent with viscosity $\eta_{sb}$ formed by the sea of blobs
and from Eqs. (\ref{ohshimalamdacero}) and (\ref{radius inner spheres})
the intrinsic viscosity of the embedded stars is

\begin{equation}
\left[\eta\right]=\frac{5}{2}\left[\frac{a}{b}k_{2}^{-1}\left(\frac{c}{c*}\right)^{3/4}\right]^{3}.\label{viscintsd}
\end{equation}
The sea of blobs exerts a strong opposition to flow because movement
proceeds through a disentanglement of the stars from their neighbors
\cite{erwin} and since they consist of a semi-dilute polymer suspension
we suggest to model the viscosity of the sea of blobs using the scaling
relation

\begin{equation}
\eta_{sb}=\eta_{sb}^{0}\left(\frac{c}{c^{*}}\right)^{\gamma},\label{viscosity sb}
\end{equation}
where $\eta_{sb}^{0}$ and $\gamma$ are additional fitting constants
that allow to model the large viscosity of the sea of blobs.

Finally, substituting Eqs. (\ref{phistar}) and (\ref{viscintsd})
and replacing the viscosity of the solvent $\eta_{0}$ in Eq. (\ref{viscosity2})
with the viscosity of the sea of blobs $\eta_{sb}$, we obtain the
viscosity for $c$ above $c^{*}$.

Using this approach we can reproduce the viscosity-concentration curve
of the star with $f=32$ above $c^{*}$ as shown in Fig. \ref{roovers1}(c).
Notice that in this regime, the volume fraction occupied by the stars
decreases with concentration and this effect alone would produce a
decrease of the viscosity with concentration. However, this decrease
is offset by the increase of the viscosity of the sea of blobs with
concentration. The parameters $\alpha$ and $\beta$ used in this
region are the ones obtained in the dilute regime and $\eta_{sb}^{0}$
and $\gamma$ are also shown in Table \ref{tabla 2}. Finally, in
panel (d) we show the case of a diblock copolymer star with $64$
arms, an inner core composed of polybutadiene arms with $N_{inner}=20000$
and an outer shell of polystyrene arms with $N_{outer}=10000$. The
viscosity data were collected in toluene which is a good solvent for
both blocks. Again, two different regions are apparent and additionally
a shoulder appears at a concentration near $c^{*}$. Surprisingly,
our model reproduces correctly this shoulder. Therefore, our model
suggests that the origin of this feature is an intrinsic feature of
the softness of the particles. The semi-dilute region is again well
described by the Daoud-Cotton picture. Notice that in all the cases
considered the thickness of the porous layer is very small as compared
to the core size (see the values of the parameter $a/b$ in Table
\ref{tabla 2}). In other words, for any practical purpose, the star
polymers behave as non porous particles and the value $\left[\eta\right]=5/2$
can be used.

Summarizing, in this work we have presented a simple model for the
calculation of the concentration dependence of the viscosity for suspensions
of soft spherical particles consisting of a hard core of radius $a$
covered with a porous shell of thickness $\delta=b-a$. The model
contains three main ingredients, the first one consists in using an
accurate model for the viscosity originally developed for rigid particles,
Eq. (\ref{viscosity2}). Secondly, we use a simple expression for
the intrinsic viscosity of permeable core-shell particles, Eq. (\ref{ohshimalamdacero})
to allow the possibility that the outer region of the particle is
solvent permeable. Third and most importantly, we incorporated in
a very simple way the effects of the softness of the particles through
Eq. (\ref{softness}). Using this model we were able to reproduce
the experimental results for the viscosity of star-branched polymers
and explained the appearance of a shoulder seen in the viscosity concentration
curves of the diblock star near the overlap concentration. We conclude
that this peculiar behavior is an intrinsic feature of the softness
of the particles and there is not need to invoke alternative mechanisms
as a shrinking of the polybutadiene block \cite{roovers} when the
total polymer concentration increases. Additionally, in the semi-dilute
regime $c/c^{*}>1$, the suspension is modeled using the Daoud-Cotton
picture which completes the description of the viscosity in the whole
concentration regime. We have tested our model comparing with experimental
results finding an excellent agreement in spite of the numerous simplifications
used in the obtention of our model.

{\LARGE Acknowledgements}{\LARGE \par}

This work was supported in part by Grant DGAPA IN-115010.

\begin{table}
\centering{}%
\begin{tabular}{|c||c|c|c|c|}
\hline 
System  & 128 arms & 64 arms & 32 arms & Diblock star 64 arms\tabularnewline
\hline 
\hline 
$a/b$ & $0.988$ & $0.989$ & $0.993$ & $0.984$\tabularnewline
\hline 
$\alpha$ & $1.90$ & $1.60$ & $1.74$ & $1.50$\tabularnewline
\hline 
$\beta$ & $0.244$ & $0.580$ & $0.966$ & $0.662$\tabularnewline
\hline 
$\eta_{sb}^{0}$ & - & - & $9.28$ & $17.4$\tabularnewline
\hline 
$\gamma$ & - & - & $2.67$ & $2.88$\tabularnewline
\hline 
\end{tabular}\caption{Fitting parameters of the model for the systems of Fig. \ref{roovers1}.}
\label{tabla 2}
\end{table}
\newpage{}

{\LARGE Figure Captions}\bigskip{}

Fig. 1. Schematics of a model for a suspension of core-shell spheres.
A hard core of radius $a$ covered by a porous layer of thickness
$\delta$. The total radius of the particle is $b=a+\delta$.

Fig. 2. Relative viscosity $\eta(\phi)/\eta_{0}$ as a function of
the particle volume fraction $\phi$ for star polymers with (a) $128$,
(b) $64$, and (c) $32$ arms and (d) diblock copolymer micelle with
$64$ arms in the good solvent toluene. Experimental data taken from
Ref. \cite{roovers}.

Fig. 3. Daoud and Cotton model for a semi-dilute suspension of star
polymers.

\newpage{}

\end{document}